\documentclass[aps,groupedaddress,a4paper,floatfix,twocolumn,footinbib,showpacs]{revtex4}
\usepackage{pslatex}
\usepackage{amsfonts}
\usepackage{amssymb}
\usepackage{mathrsfs}
\usepackage[mathscr]{euscript}
\usepackage{graphicx}
\usepackage{fancyhdr}
\usepackage[hypertex=true]{hyperref}

\begin{document}

\title{Wideband pulse propagation: single-field and multi-field
approaches to Raman interactions}
\author{P. Kinsler}
\affiliation{
  Department of Physics, Imperial College London,
  Prince Consort Road,
  London SW7 2BW, 
  United Kingdom.
}
\author{G.H.C. New}
\affiliation{
  Department of Physics, Imperial College London,
  Prince Consort Road,
  London SW7 2BW, 
  United Kingdom.
}

\lhead{
\includegraphics[height=5mm,angle=0]{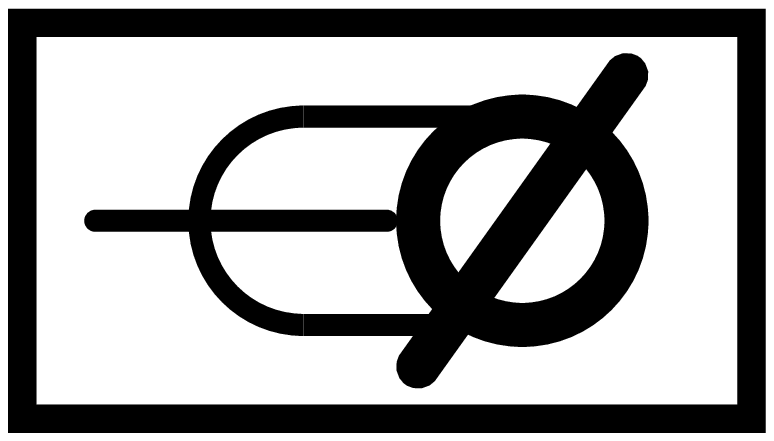}~~
{ WBRAM }}
\chead{{\em Wideband pulse propagation: ...}}
\rhead{
\href{mailto:Dr.Paul.Kinsler@physics.org}{Dr.Paul.Kinsler@physics.org}\\
\href{http://www.qols.ph.ic.ac.uk/}{http://www.qols.ph.ic.ac.uk/$\sim$kinsle/}
%http://www.kinsler.org/physics/}
}
%\lfoot{X}

\date{\today}

\begin{abstract}

We model the process of ultra broadband light generation in which a pair of 
laser pulses separated by the Raman frequency drive a Raman transition.
In contrast to the usual approach using separate field envelopes for the 
different frequency components, we treat the field as a single entity.
This requires the inclusion of few-cycle corrections to the pulse propagation.
Our single-field model makes fewer approximations and is mathematically 
(and hence computationally) simpler, 
although it does require greater computational resources to implement.  
The single-field theory reduces to the traditional multi-field 
one using appropriate approximations.

\end{abstract}


\maketitle
\thispagestyle{fancy}

% --------- --------- --------- --------- --------- --------- ---------

{\em
Published as
Phys. Rev. A{\bf 72}, 033804 (2005).
A detailed derivation of the single-field theory
can be found at http://arxiv.org/abs/physics/0606112
}

\section{Introduction}\label{S-Intro}

An important aim of current wideband Raman experiments
is to  efficiently generate few-cycle pulses
\cite{Harris-S-1998prl,Sokolov-WYYH-2001prl,Hakuta-SKL-2000prl,Sali-MTHM-2004ol}.  
If driven strongly enough, the two-photon Raman transition
modulates the incoming field by adding sidebands separated by the transition
frequency.  Wideband fields are generated as these sidebands generate
sidebands of their own (and so on); a wide comb of frequency
components separated by the transition frequency is generated in this way. 
If a scheme can be implemented that adjusts the phases of 
each component appropriately, 
then few- or single- cycle optical pulses can be obtained 
(see e.g. \cite{Sokolov-WYYH-2001prl}).  

In standard theoretical treatments of the Raman process,
the field is split into frequency components 
centred on the teeth of the frequency comb.
This approach has the advantage that the components can be 
modeled reasonably well with slowly varying envelopes, 
but of course it has the disadvantage that one needs to keep track of 
a large number of components.  
In this paper, 
we study an alternative approach in which the field 
is treated as a single entity rather than being split into pieces.  
Note that this approach is
distinct from methods based on the direct solution of 
Maxwell's equations such as 
FDTD (finite difference time domain)\cite{Joseph-T-1997itap}
or PSSD (pseudospectral spatial domain)\cite{Tyyrell-KN-2005jmo}.
Our single-field is based on a second-order wave equation, 
and uses a convenient choice of carrier to define a field envelope.  
As we will demonstrate, 
the latter technique offers significant advantages over 
the traditional multi-field formalism. 

To provide a context for the discussion, 
  we consider experiments such as those of 
Sali et.al. \cite{Sali-MTHM-2004ol,Sali-KNMHTM-2004draft}, 
where the Raman transition is driven near-resonantly by 
  a pair of intense pump pulses about 100fs long; 
compared to the transition frequency of about 130THz, 
  the spectra of each pump pulse (and hence the generated sidebands) 
  are relatively narrow.  
This means that a multi-component model is still not unreasonable, 
  even if numerical considerations might demand that 
  the arrays used to store these spectra overlap in frequency space.  
However, if we were to move to shorter pump pulses, 
  or to a single (much shorter) pump pulse with enough bandwidth to
efficiently excite the transition,
   we would reach the regime where the
   teeth of the frequency comb significantly overlap.  
At this point,
  one would be forced not only to move from a solution based on the
Slowly-Varying Envelope Approximation (SVEA) to a more accurate model
such as the Generalized Few-Cycle Envelope Approximation (GFEA) 
\cite{Kinsler-N-2003pra,Kinsler-FCPP},
but the utility of multiple field components would in any case
become questionable. 
 This provides the motivation for the present work, 
since our model can extend the regime in 
      which the advantages of envelope-based methods can be utilised;
it also turns out to be more versatile, 
placing fewer restrictions on the kinds of 
Raman media that can be easily described.

%In this regime it can be advantageous to treat the field as a single unit,
%rather than splitting it into pieces.  

In this paper, 
we will construct a single-field model which, in most other respects, 
closely parallels the approach to wideband Raman
generation adopted by Hickman et. al. \cite{Hickman-PB-1986pra}.  
A key feature of the single-field model is that 
  the coupling constants oscillate at the Raman frequency, 
  and it is this that impresses the sideband modulation on the
  propagating field.  
Since the field is now not only wide-band but
  contains significant sideband components 
  (i.e. distinct sub-peaks, as opposed to a broad featureless background),
the field envelope is no longer slowly-varying and must therefore 
  be propagated using the GFEA.  
This necessity can be demonstrated by comparing 
  the results of the single-field model with those of 
  a multi-field counterpart.

The paper is organized as follows: 
%following this introduction (section \ref{S-Intro}), 
section \ref{S-Theory} outlines the derivation of 
  the single-field Raman theory,
section \ref{S-Multi} shows how to reduce it to a standard multi-field version,
and section \ref{S-Application} applies the theory to
practical situations.  
In section \ref{S-Discuss} we discuss some of the issues relating
to our Raman model and its numerical implementation, and 
finally section \ref{S-Conclude} contains our
conclusions.

% --------- --------- --------- --------- --------- --------- ---------

\section{Single-field Raman theory}\label{S-Theory}

We start by considering the wave function
$\psi$ of a single molecule (e.g. H$_2$) and the electric field $E$, and
write the time-dependent wave function by expanding it in terms of the
eigenfunctions in the field-free (i.e. $E=0$) case.  
This means we can get the expansion
coefficients by solving for an effective Schr\"odinger equation that contains
a two-photon Rabi frequency created by means of an interaction term based on a
field-dependent dipole moment.  We assume a dispersionless medium and write all
equations in terms of position $z$ and retarded times $t=t_{lab}-z/c$.  
Here we follow the method of Hickman, Paisner, and Bischel
\cite{Hickman-PB-1986pra} (HPB), 
but we use only a single $E$ field rather than multiple components.  
Note that HPB use {\em Gaussian} units, so there may appear to be 
inconsistencies when comparing our formulae (in S.I.) to theirs.

We denote the known molecular eigenfunctions of the 
unperturbed Hamiltonian $H_0$ as $\left| n \right>$, and their
corresponding energies $\hbar W_n$.  We want to 
obtain the solution to 
~
\begin{eqnarray}
\left( H_0 + V \right) \psi 
&=& 
  \imath \hbar 
  \frac{\partial \psi}
       {\partial t}
,
\label{eqn-hamltonian-def}
\\
\textrm{with} ~~~ ~~~
V &=& 
-d E 
,
\label{eqn-perturbation-def}
\\
\psi &=&
\sum_n
  c_n e^{-\imath W_n t} \left| n \right>
,
\label{eqn-psi-def}
\end{eqnarray}
~
where $d$ is the electronic dipole moment operator and the $c_n$ are a 
set of complex probability amplitudes.

We now replace the electric 
field $E$ with a carrier-envelope 
description, but, unlike HPB, we use only a single 
component 
centred at a frequency of $\omega_0$,
rather than a set indexed by an integer $j$.
The envelope and carrier for the field is:
~
\begin{eqnarray}
E &=& 
    A 
    e^{
     -\imath 
        \left( 
            \omega_0 t 
           -
             k_0 z 
        \right)
      } 
  + \textrm{c.c.}
,
\label{eqn-single-EfromA}
\end{eqnarray}
and, 
following the standard procedure 
of 
assuming the co-efficients $c_i$ 
are slowly varying, discarding terms at multiples of the carrier frequency,
and simplifying, 
 we eventually reach
~
\begin{eqnarray}
\imath \hbar \frac{d c_n}{dt}
&=&
-
\sum_j
  c_j
  \alpha_{nj}
   ~~ . 2 \left| A \right|^2
,
\label{eqn-single-A-DcnDt}
\\
\textrm{where} ~~~~
\alpha_{nj}
&=&
      \frac{1}{ \hbar}
          \exp \left[ -\imath W_{jn} t \right]
  \sum_i
      d_{ni} 
      d_{ij} 
    \frac{W_{ij} }
         {W_{ij}^2-\omega_0^2}
.
\label{eqn-single-A-alpha}
\end{eqnarray}

The $\alpha_{nj}$ coupling parameters oscillate because,
 in contrast to the HPB derivation, there is no frequency difference between 
field components to cancel with the Raman transition frequency.  
We now take the indices $1$ and $2$ to correspond to the two states 
involved in the Raman transition of interest; 
these will be the $0$ and $1$ vibrational (or perhaps rotational) levels 
of the electronic ground state.  
Indices $3$ and above will correspond to (quoting HPB) 
``translational motion on higher electronic states''.  
Since we are interested only in the Raman transition, 
we specialize the above equations for the coefficients $c_n$, 
calculating $c_1$ and $c_2$ only, 
and assuming that the $d_{12} = \left< 1 \right| d \left| 2 \right>$ 
dipole moment is zero.  
This means we will only be including transitions between indices $1$
and $2$ that {\em go via one of the higher states} $j \ge 3$, since we still
allow $d_{1j}, ~d_{2j} \neq 0 ~~ \forall j \ge 3$.  
Further, we solve for the
coefficients for the higher states in terms of $c_1$ and $c_2$, in 
an adiabatic approximation justified when $c_1$ and $c_2$ vary only slowly
compared to the exponential terms.

When converting the equations for $c_1$, $c_2$ into Bloch 
equations, we make the same approximations as HPB:
keeping the energy separations 
for all transitions greater than that of the $1 \leftrightarrow 2$ transition,
and ignoring all the higher vibrational (or rotational)
states.  
Thus we can write 
~
\begin{eqnarray}
  \alpha_{12}^* - \alpha_{21}
&\approx&
  0,
\\
  \alpha_{12} + \alpha_{21}^*
&=&
  2 \hbar f' e^{-\imath \omega_b t + \imath \delta'}
.
  \label{eqn-alpha-bar}
\end{eqnarray}
  Here $\omega_b$ is the Raman transition frequency, 
  and $\delta'$ is a phase factor that ensures 
  that the coupling constant $f'$ is real valued. 
  This $f'$ will be used to 
  replace $\alpha_{12}+\alpha_{21}^*$.
We also get a Stark shift term -- 
~
\begin{eqnarray}
  \hbar g'
&=&
  \alpha_{11}' - \alpha_{22}'^*
.
  \label{eqn-alpha-stark}
\end{eqnarray}

We define $\rho_{12}=c_1 c_2^*$ and 
$w=c_2^* c_2 - c_1^* c_1$, so that 
~
\begin{eqnarray}
\frac{d \rho_{12}}{dt}
&=&
  \imath 
  \frac{\left( \alpha_{11} - \alpha_{22}^* \right) }
     {\hbar}
  2 \left| A \right|^2
  \rho_{12}
+ \imath 
  \frac{\alpha_{12} }{ \hbar}
  2 \left| A \right|^2
  w
,
\label{eqn-basic2pbloch-rho}
\\
\frac{dw}{dt}
&=&
+ \imath
  \frac{2 \alpha_{12}^* }
       { \hbar}
  2 \left| A \right|^2
    \rho_{12}
   - 
 \imath
  \frac{ 2 \alpha_{12} }
       { \hbar}
  2 \left| A \right|^2
    \rho_{12}^*
.
\label{eqn-basic2pbloch-w}
\end{eqnarray}

Finally, we insert decay terms $\gamma_i$,
and introduce $\omega_b'=\omega_b-\Delta$.  
This $\Delta$ allows for arbitrary rotations of the polarization, 
$\rho_{12} 
=
\rho_{12}' \exp \left( -\imath \Delta t -\imath \delta' \right)
$.
Eqns. (\ref{eqn-basic2pbloch-rho},\ref{eqn-basic2pbloch-w})
governing the response of the medium to the applied fields now become
~
\begin{eqnarray}
  \partial_t \rho_{12}' 
&=& 
  \left(
   -\gamma_2 
   + \imath \Delta 
  \right)
  \rho_{12}' 
\nonumber
\\
& &
~~~~ ~~~~
+ 
  \imath g' 2 A^* A 
  \rho_{12}' 
  ~
+ 
  \imath f' 
  2 A^* A w  
  e^{ \imath \omega_b' t }
,
\label{eqn-rbpostRWA-last-rho}
\\
\partial_t w
&=&
  - \gamma_1 \left( w' - w_i \right)  
\nonumber
\\
& &
~~~~ ~~~~
  +
   2 \imath  f' 
   .
   2 A^* A 
   \left( 
      \rho_{12}' 
      e^{ \imath \omega_b' t }
     -
      \rho_{12}'^* 
      e^{-\imath \omega_b' t }
   \right)
.
\label{eqn-rbpostRWA-last-w}
\end{eqnarray}
The parameter $\Delta$ should be chosen to optimise computational accuracy by
making the dynamics as slowly-varying as possible.  
For example, if the field contained two frequency components that 
were slightly detuned from the Raman frequency, 
we might use $\Delta$ to compensate for the resultant beating.  
In general, 
$\Delta$ is most useful in the multi-field model discussed in the next section.
The complementary part that specifies how the field responds to the 
polarization of the Raman transition, is
~
\begin{eqnarray}
  \partial_z A(t) 
&=&
  \frac{2 \imath \pi \omega_0}{c_0 n_0} 
  \times
    \left[
       1 + \frac{\imath \partial_t }{\omega_0}
    \right]
  \frac{ \mathscr{B}(t) }{4 \pi \epsilon_0}
\\
&=&
  \imath 
  \frac{2 \sigma \bar{\alpha}_{12} \omega_0}{c_0 n_0 \epsilon_0 } 
  \times
    \left[
       1 + \frac{\imath \partial_t }{\omega_0}
    \right]
  A(t) X(t)
,
  \label{eqn-single-Apropagate}
\\
  X(t)
&=& 
%  \left[
      \rho_{12}' 
      e^{ \imath \omega_b' t }
     +
      \rho_{12}'^* 
      e^{-\imath \omega_b' t }
%  \right]
  \label{eqn-single-X}
.
\end{eqnarray}

Here the $1 + \imath \partial_t/\omega_0$ in eqn.(\ref{eqn-single-Apropagate})
is (with $\partial_t \equiv d/dt$) the lowest-order approximation to the
GFEA few-cycle propagation corrections 
\cite{Kinsler-FCPP,Kinsler-N-2003pra},
which is equivalent to the SEWA (Slowly Evolving Wave Approximation) 
correction derived by Brabec and Krausz \cite{Brabec-K-1997prl}. 
Although the full form is not included  for reasons of brevity,
it could easily be introduced if the extra accuracy was desired; 
indeed we routinely use it in our simulation codes.  
It is 
independent of the Raman derivation presented here, since 
it is a field propagation effect. The full form of the few-cycle 
prefactor (and various expansions thereof) has already been reported in 
\cite{Kinsler-N-2003pra,Kinsler-FCPP}.

A detailed derivation of this single-field Raman theory can be found in 
\cite{Kinsler-2006arXiv-sfwbr}

We solve these equations numerically using a split step method, 
where we treat the nonlinearity in the time domain, 
and the dispersion in the frequency domain.
To include dispersion in a time domain equation like 
eqn.(\ref{eqn-single-Apropagate}) requires either additional time derivatives
(as in \cite{Kinsler-FCPP,Kinsler-N-2003pra}) 
or a convolution over a time-response function which is an $N^2$ operation.  
However, 
handling dispersion in the frequency domain is both conceptually simpler 
(since it simply amounts to a frequency-dependent phase evolution), 
and more computationally efficient because it is an $N \log N$ process.

  The validity of the approximations used in deriving our Bloch equations 
will obviously depend both 
on the details of the chosen Raman medium and/or transition, and on the
number of Stokes and anti-Stokes sidebands we wish to describe.  Since in the 
experiments of \cite{Harris-S-1998prl,Sokolov-WYYH-2001prl,Hakuta-SKL-2000prl,Sali-MTHM-2004ol,Gundry-AASTKNM-2005ol} the emphasis was on a single
Raman transition, a simple Bloch model is clearly appropriate, and indeed 
our approximations differ little from those of other theoretical 
approaches (such at that of HPB).

% --------- ---------

\section{Multi-field Raman Theory}\label{S-Multi}

The single-field Raman model  can be converted into a  traditional
multi-field model as developed in e.g.  HPB \cite{Hickman-PB-1986pra} or Syed,
McDonald and New 
\cite{Syed-MN-2000josab} by  replacing the field envelope with a sum of
multiple envelopes using carrier exponentials spaced at the Raman frequency.
When doing this, we will only get the correct multi-field form if few-cycle 
(either SEWA or GFEA) corrections to the field evolution part of
the theory are applied to the effective polarization caused by the 
Raman transition.

Since the single-field evolution equation (eqn.(\ref{eqn-single-Apropagate})) 
uses an envelope $A$ that is based on a carrier
(see eqn.(\ref{eqn-single-EfromA})), 
the single-field envelope $A$ is replaced with $A_j$'s 
at frequency $\omega_j = \omega_0 + j \omega_b$ 
and wavevector $k_j = k(\omega_j)$.  
The single-field envelope in terms of the new $A_j$'s is
~
\begin{eqnarray}
  A 
&=&
  \sum_j 
    A_j 
    \exp 
      \left[
        -\imath 
        \left( 
            \omega'_j t 
           -
             k'_j z 
        \right)
      \right]
,
\label{eqn-multienvelope}
\end{eqnarray}
where $\omega'_j = \omega_j - \omega_0$, and $k'_j = k_j - k(\omega_0)
= k_j - k_0$.

The equations 
for $ \rho_{12}'$ and $w$ describing the Raman transition 
 result from a simple substitution 
of eqn.(\ref{eqn-multienvelope}) 
into eqns.(\ref{eqn-rbpostRWA-last-rho}, \ref{eqn-rbpostRWA-last-w}), 
followed by a rotating wave approximation (RWA) 
to remove non frequency matched terms. 
They are
~
\begin{eqnarray}
  \partial_t \rho_{12}' 
&\approx&
  \left(
     -\gamma_2 + \imath \Delta + \imath g' \sum_j 2 A_j^* A_j
  \right) 
  \rho_{12}'
\nonumber
\\
&& ~~~~ ~~~~
+ 
2 \imath f' \sum_j 2 A_{j}  A_{j-1}^*
. w 
. e^{-\imath \Delta t } 
. e^{+\imath \left( k_j-k_{j-1}  \right) z }
,
\label{eqn-multi-dr}
\\
\partial_t w
&=&
- \gamma_1 \left( w - w_i \right)  
,
\nonumber
\\
&&
+
   2 \imath f' 
   \left( 
      2 A_j^*A_{j+1}
      \rho_{12}' 
      e^{ \imath \omega_b' t }
     -
      2 A_j A_{j+1}^*
      \rho_{12}'^* 
      e^{-\imath \omega_b' t }
   \right)
.
~~~~ ~~~~
\label{eqn-multi-dw}
\end{eqnarray}

Quite a lot of physics has been removed by the RWA approximation, 
although it is a very reasonable one except in the very wideband limit.
For example, the effects of next-nearest neighbour field components
%acting on the transition 
have been ignored, 
as have all more distant field-field interactions. 
In the next-nearest neighbour case, the dropped terms would impose a 
rapid $\omega_b$ oscillation onto the polarization $\rho_{12}$, 
which would in turn tend to impose sidebands at $\pm \omega_b$ onto
each field component.  
It is reasonable to ignore such sidebands
in the narrowband limit used for most applications of a multi-field Raman 
theory; 
but, in principle one might extend a multi-field theory to include them 
by inventing a scheme to apply the sidebands to the field component 
with which they are in nearest resonance.

Extra factors of $2$ have appeared in eqns.(\ref{eqn-multi-dr}, 
\ref{eqn-multi-dw}) 
because the multi-field equations start with double summations that give 
pairs of terms that can be reduced to one in the remaining single summation.  

Finally, we need to insert the few-cycle correction 
to the polarization term, because the ($j\ne 0$) sub-envelopes $A_j$
have an $\imath j \omega_b t$  time dependence that cannot be neglected.
The polarization correction terms are just the result of
applying the first-order correction $(\imath/\omega_0)\partial_t$ to
the $A(t)X(t)$ from eqn.(\ref{eqn-single-Apropagate}).
The $j$-th polarization correction term is then
~
\begin{eqnarray}
&&
  \imath
  \frac{ \sigma \omega_j' \alpha_{12}}
       {2 \epsilon_0 c_0}
  \left\{
      \rho_{12}'
      A_{j+1}
      \exp\left[ +\imath (k'_{j+1} -k'_j) z - \imath \Delta t \right]
\right.
\nonumber
\\
&& ~~~~ ~~~~ ~~~~ ~~~~
\left.
  +
      \rho_{12}'^*   
      A_{j-1}
      \exp\left[ +\imath (k'_{j-1} -k'_j) z + \imath \Delta t \right]
\right\}
\nonumber
\\
&& ~~~~ ~~~~ ~~~~ ~~~~ 
 - 
  \imath \left( k_j - k_0 \right) A_j
,
\end{eqnarray}
and differs only from the standard polarization term in that $\omega_j'$ 
appears in place of $\omega_0$.  The two terms can then 
be straightforwardly summed, and since $\omega_j=\omega_0+\omega_j'$,
from
eqns.(\ref{eqn-single-Apropagate}, \ref{eqn-single-X}, 
\ref{eqn-multienvelope}), we get
~
\begin{eqnarray}
  \partial_z A_j(t)
&=&
  \imath
  \frac{ \sigma \omega_j \alpha_{12}}
       {2 \epsilon_0 c_0}
  \left\{
      \rho_{12}'
      A_{j+1}
      \exp\left[ +\imath (k'_{j+1} -k'_j) z - \imath \Delta t \right]
\right.
\nonumber
\\
&& ~~~~ ~~~~ ~~~~ ~~~~
\left.
  +
      \rho_{12}'^*   
      A_{j-1}
      \exp\left[ +\imath (k'_{j-1} -k'_j) z + \imath \Delta t \right]
\right\}
\nonumber
\\
&& ~~~~ ~~~~ ~~~~ ~~~~ 
 - 
  \imath \left( k_j - k_0 \right) A_j
,
\label{eqn-multi-Ajpropagate}
\end{eqnarray}
where the $\imath \Delta t$ terms arise because of our
rotation of the frame of reference of $\rho_{12}'$. The residual 
$k_j - k_0$ terms result from a difference in the $k$ frame
of reference between the our multi-field derivation and the standard one.

%% A detailed derivation of this can be found in 
%% \cite{Kinsler-2005arXiv-sfwbr}

% --------- --------- --------- --------- --------- --------- ---------

\section{Example Applications}\label{S-Application}

We now use the single-field (GFEA) model to simulate an experimental 
situation.  
First we compare the results to their multi-field counterparts,
demonstrating the relationships between the two methods, 
and showing them to be in good agreement, as expected for the chosen
pulse lengths.
Second, we contrast our model with an (inaccurate) single-field SVEA model, 
in order to highlight the role of the few-cycle propagation terms.
The bulk of the code used was the same for all simulations, 
as it contains options to switch from a single to a multi-field case,
and to switch GFEA corrections on and off. 

\begin{figure}
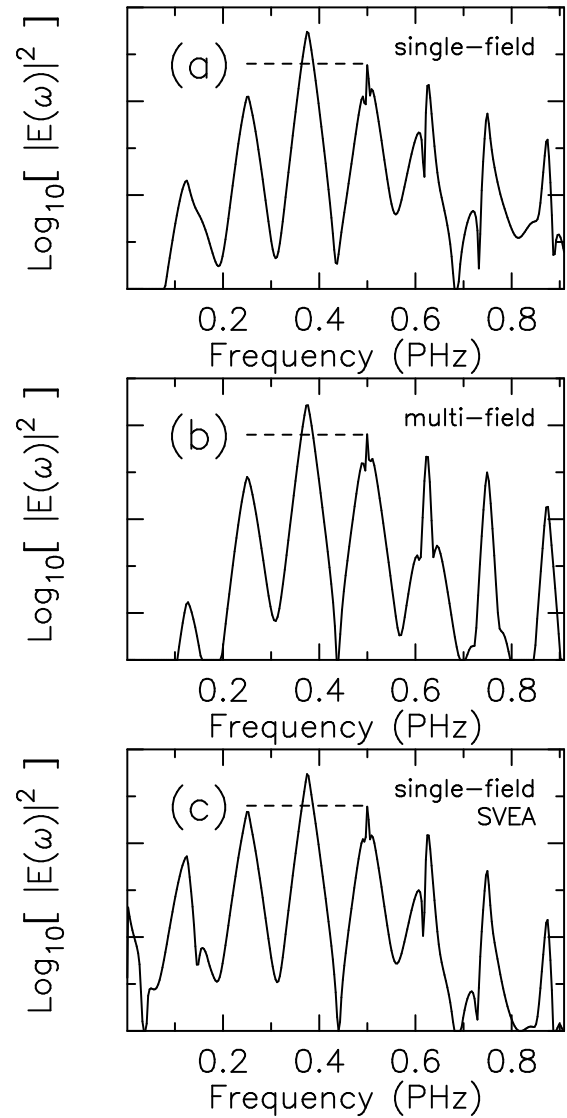
 % ?? [h]
\includegraphics[width=49mm,angle=-90]{f01a-tr-sf-GFEA.ps}\\
\includegraphics[width=49mm,angle=-90]{f01b-tr-mf-GFEA.ps}\\
\includegraphics[width=49mm,angle=-90]{f01c-tr-sf-SVEA.ps}
\caption{
Transient Raman generation using 17.5fs and 62.5f pump pulses as described
in the text. Here we compares three simulation results:
(a) single-field GFEA simulation, 
(b) multi-field simulation, and 
(c) single-field SVEA simulation. 
The dashed lines help compare the relative heights of the first 
Stokes and anti-Stokes peaks.  
The vertical scale is in arbitrary units.
}
\label{F-transient-pump}
\end{figure}

Figure \ref{F-transient-pump} shows a set of results for a pair of 
pump pulses traveling though 9cm of H$_2$.  
This corresponds to a simulation of an experiment where
the pulses pump the 1st vib(ro) level in molecular H$_2$ (at 4155cm$^{-1}$, 
i.e. $\sim 126$THz), 
as in the transient-regime experiments of 
Sali et.al. \cite{Sali-MTHM-2004ol,Sali-KNMHTM-2004draft}.  
In these experiments, typical pulses might be 70fs and 250fs wide
at 800nm (30$\mu$J) and 600nm (120$\mu$J) respectively, 
and the comb of Raman sidebands generated 
are narrow and well separated.  
A Cauchy-type dispersion curve for H$_2$ is incorporated into the simulations.
In our simulations, 
we use the smaller widths of 17.5fs and 62.5fs, which broadens 
the spectral peaks (to about 57THz and 16THz respectively) 
and makes the standard multi-field approach less practical.
The figure compares three  data sets -- 
(a) single-field GFEA simulation, 
(b) multi-field simulation, 
and lastly (c) single-field SVEA simulation (i.e. {\em
without} any few-cycle propagation corrections).  

There is good agreement in the heights of all the spectral peaks between the 
two exact simulations (single-field GFEA fig. \ref{F-transient-pump}(a) and 
multi-field fig \ref{F-transient-pump}(b) ); even the 
details in the wings of the first anti-Stokes peak (at about $f=0.5$)
are replicated.  Those in the wings of the second 
anti-Stokes peak (at about $f=0.63$) are not well replicated;
however, the features in question are about three orders of magnitude 
weaker than the peaks, and the two simulations are not 
equivalent because the multi-field theory does not include
next-nearest neighbour interactions.

The comparison between fig. \ref{F-transient-pump}(a,b) and the 
single-field SVEA simulation fig \ref{F-transient-pump}(c) is also 
instructive.  Although it does reproduce the character of the single-field 
GFEA spectra in many ways, the peak heights do not agree -- a fact 
that is more apparent on a linear scale than a logarithmic one.
In terms of a multi-field model, we can say that without the GFEA
corrections, the prefactor of the polarization term does not pick up
its correct frequency dependence, 
so the Stokes lines are artificially enhanced, 
and the anti-Stokes artificially suppressed.

% \rho_{12}=0.025\imath ==> v = 0.05

\begin{figure}
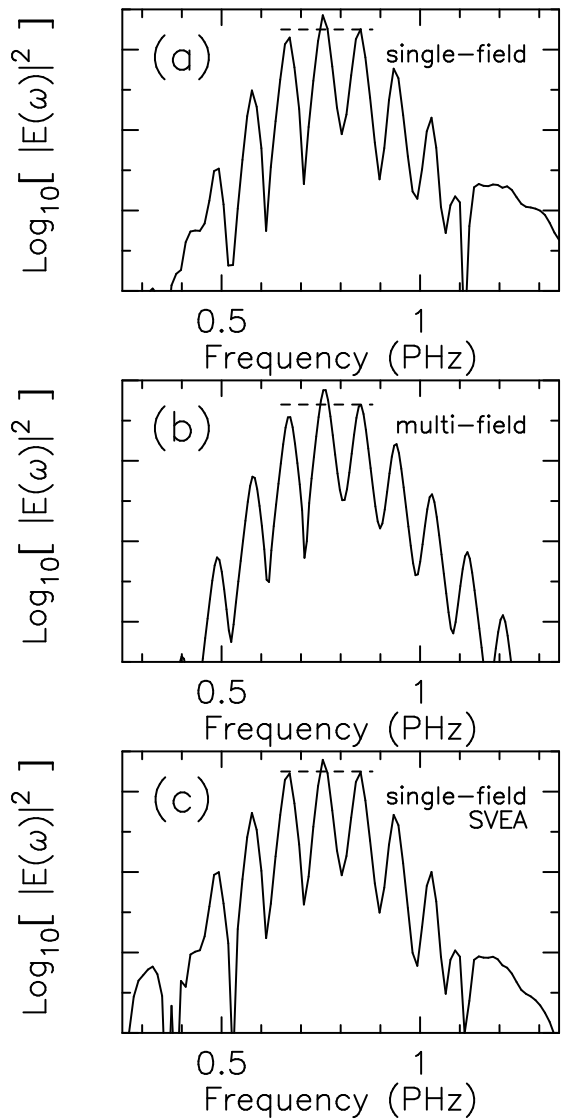
 % ?? [h]
\includegraphics[width=49mm,angle=-90]{f02a-ad-sf-GFEA.ps}\\
\includegraphics[width=49mm,angle=-90]{f02b-ad-mf-GFEA.ps}\\
\includegraphics[width=49mm,angle=-90]{f02c-ad-sf-SVEA.ps}
\caption{
10fs probe pulse incident on a medium with an initial 
polarization of $\rho_{12}=0.025\imath$.  
Here we compares three simulation results:
(a) single-field GFEA simulation, 
(b) multi-field simulation, and 
(c) single-field SVEA simulation. 
The dashed lines help compare the relative heights of the first 
Stokes and anti-Stokes peaks.
The vertical scale is in arbitrary units.
}
\label{F-adiabatic-probe}
\end{figure}

Figure \ref{F-adiabatic-probe} shows a set of results from a single 
10fs probe pulse at 397nm,
traveling though 9cm of previously polarized D$_2$.
This corresponds to the probe stage of an experiment where
the gas had been prepared using a pair of nanosecond
fields resulting in a medium polarization of $\rho_{12}=0.025\imath$ on the 
2993.57cm$^{-1}$ ($\sim$ 90THz) vibrational transition,
e.g. as in the 
experiments of Gundry et.al. \cite{Gundry-AASTKNM-2005ol}, 
who use a longer 
probe pulse of about 150fs.  
A Cauchy-type dispersion curve is incorporated into the 
simulations, but in the absence of 
good dispersion data for D$_2$, we use that for $H_2$ 
as it should be a good match.
      Note that although the polarization initial condition is fixed, 
      our simulations do incorporate the response of the polarization to
      the probe pulses.
The main spectral peaks agree well 
in the multi-field and single-field GFEA simulations, 
although as before the results differ at the edges 
where the intensities are very small compared to the main features.  
As for the previous situation, 
in the single-field SVEA simulation the Stokes and anti-Stokes lines are 
artificially enhanced or suppressed.

% --------- --------- --------- --------- --------- --------- ---------
\section{Discussion}\label{S-Discuss}

For simple systems, 
those (for example) with a single Raman transition driven by 
relatively long pulses, 
it will usually be most efficient to continue using a multi-field model.
Single-field simulations require very fine time-resolution,
so they are computationally expensive for 
pulses with many optical cycles.
The spectral range of the numerical field is correspondingly broad,
typically covering many Stokes and anti-Stokes lines.

In more complex situations, however,
the single-field approach will outperform its multi-field counterpart.
For example, 
if a Raman interaction is probed by a beam that does not lie on the frequency 
comb defined by the pumping beams (e.g. as in \cite{Gundry-AASTKNM-2005ol}),
the multi-field approach will become much more complicated to implement.
It will be necessary to define separate arrays for the 
pump and probe Raman ``ladders'' of Stokes and anti-Stokes lines,
an issue that we avoided in section \ref{S-Application}
by replacing the pump stage of the process with an 
initial condition for the polarization.  
With a single-field model, 
the probe pulse and its Raman sidebands simply get superimposed on the 
overall spectrum, where they will be offset from the frequency ladder defined
by the pump beams.

Another situation in which the multi-field model will run into difficulty is
where there are multiple Raman resonances.
Although the treatment in this paper has been restricted to a simple two-level
      Bloch equation description of the Raman medium,
additional Bloch equations can easily be added,
even if there are coupled multi-level interactions
(as for example in \cite{Wallis-1995pra}).
It is only necessary to describe those transitions appropriately, 
and to modify the polarization terms acting on the propagating field.
This procedure is considerably more difficult to handle in the 
multi-field case,
which is based on field components separated by a 
particular Raman transition frequency.
Additional  Raman resonances  complicate the theory; 
not only must extra detuning factors be added to the equations, 
but it is also necessary to work out which field component is nearest to
each new driving term.
With a wideband single-field model, on the other hand,
any new sidebands or resonance effects appear automatically
in the spectrum, and no special measures need to be adopted
to handle them.

The usefulness of our single-field approach is not restricted to the 
Raman interaction described in this paper.  
It is not just more easily extended to more complex Raman 
      materials involving e.g. multiple transitions than the standard 
      multi-field model.  
It would be equally valuable for a near-degenerate optical parametric 
oscillator, or indeed any system where two or more field 
components start to overlap as the pump or probe 
pulses get shorter.

% --------- --------- --------- --------- --------- --------- ---------
\section{Conclusion}\label{S-Conclude}

We have considered how best to model the multi-frequency field 
in wideband Raman generation experiments.
Rather than using multiple field envelopes, with one at each 
Stokes or anti-Stokes frequency, we instead 
use a single wideband field envelope.  
This requires that the field be propagated
taking into account wideband effects, 
as described by either the SEWA theory 
of Brabec and Krausz \cite{Brabec-K-1997prl}, 
or the more general GFEA of Kinsler and New \cite{Kinsler-N-2003pra}.

Our single-field approach has three crucial advantages.
First, it includes more physics, 
even compared to a multi-field approach enhanced by adding GFEA corrections 
to the propagation of the field components.  
Secondly, it deals effortlessly with 
 the complications of overlapping spectra that occur in the multi-field case.  
Thirdly, it allows for extra Raman transitions, and other molecular 
details to be included more easily than is possible for the multi-field
model.

All of these factors ensure that our wideband single-field model not only 
extends the regime in which envelope-based methods can be utilised; 
but is also more versatile 
and places fewer restrictions on the kinds of 
Raman media that can be easily described.

% --------- --------- --------- --------- --------- --------- ---------

\end{document}